\documentclass[runningheads,a4paper]{llncs}

\usepackage{amssymb}
\usepackage{adjustbox}
\usepackage{graphicx}
\usepackage{siunitx}
\usepackage{mathtools}
\usepackage{threeparttable}
\usepackage{float}
\usepackage{booktabs}
\usepackage{tabularx}
\usepackage{makecell}
\newcolumntype{C}[1]{>{\centering\arraybackslash}p{#1}}
\newcolumntype{L}[1]{>{\raggedright\arraybackslash}p{#1}}
\usepackage{multicol}
\usepackage[T1]{fontenc}
\usepackage{csquotes}

\usepackage{url}
\newcommand{\keywords}[1]{\par\addvspace\baselineskip
\noindent\keywordname\enspace\ignorespaces#1}
\usepackage{subcaption}
\graphicspath{ 
{fig_money/} 
{fig_results/}
{fig_results/12899_6_006/}
{fig_results/15177_5_005/}
{fig_results/11933_5_015/}
{fig_results/19028_5_006/}
}

\begin{document}

\mainmatter

\title{Class Imbalance Correction for Improved Universal Lesion Detection and Tagging in CT}

\titlerunning{Class Imbalance Correction for Universal Lesion Detection \& Tagging}

% \author{*}
% \authorrunning{*}
% \institute{*}

\author{Peter D. Erickson, Tejas Sudharshan Mathai\inst{*}, Ronald M. Summers}

\authorrunning{P.D. Erickson et al.}

\institute{Imaging Biomarkers and Computer-Aided Diagnosis Laboratory, Radiology and Imaging Sciences, Clinical Center, National Institutes of Health, Bethesda, MD, USA \\
* Corresponding author, email: mathaits [at] nih dot gov}

\maketitle

% ================================================
\begin{abstract}

Radiologists routinely detect and size lesions in CT to stage cancer and assess tumor burden. To potentially aid their efforts, multiple lesion detection algorithms have been developed with a large public dataset called DeepLesion (32,735 lesions, 32,120 CT slices, 10,594 studies, 4,427 patients, 8 body part labels). However, this dataset contains missing measurements and lesion tags, and exhibits a severe imbalance in the number of lesions per label category. In this work, we utilize a limited subset of DeepLesion (6\%, 1331 lesions, 1309 slices) containing lesion annotations and body part label tags to train a VFNet model to detect lesions and tag them. We address the class imbalance by conducting three experiments: 1) Balancing data by the body part labels, 2) Balancing data by the number of lesions per patient, and 3) Balancing data by the lesion size. In contrast to a randomly sampled (unbalanced) data subset, our results indicated that balancing the body part labels always increased sensitivity for lesions $\geq$ 1cm for classes with low data quantities (Bone: 80\% vs. 46\%, Kidney: 77\% vs. 61\%, Soft Tissue: 70\% vs. 60\%, Pelvis: 83\% vs. 76\%). Similar trends were seen for three other models tested (FasterRCNN, RetinaNet, FoveaBox). Balancing data by lesion size also helped the VFNet model improve recalls for all classes in contrast to an unbalanced dataset. We also provide a structured reporting guideline for a ``Lesions'' subsection to be entered into the ``Findings'' section of a radiology report. To our knowledge, we are the first to report the class imbalance in DeepLesion, and have taken data-driven steps to address it in the context of joint lesion detection and tagging.

\end{abstract}
% ================================================

% ================================================
\section{Introduction}
\label{intro}
% ================================================

Tumor burden assessment and staging of cancer is critical for patient treatment \cite{Eisenhauer2009,Schwartz2016}. The first step towards this goal is lesion localization, which enables lesion size measurement and assessment of malignancy risk. Typically, in clinical practice, computed tomography (CT) and positron emission tomography (PET) are  preferred for lesion analysis \cite{Eisenhauer2009}. Radiologists scroll through a volume to find lesions of size $\geq$ 10mm and treat them as suspicious for metastasis \cite{Eisenhauer2009,Schwartz2016}. They also identify lesions across multiple patient visits and track their progression (growth, shrinkage, or unchanged status) based on treatment response. Lesions can have heterogeneous shapes, sizes, and appearances in CT, and this further compounds assessment as there are a variety of imaging scanners and inconsistent exam protocols in use at different institutions. Moreover, sizing lesions during a busy clinical day is cumbersome for a radiologist due to observer measurement variabilities, especially when treatment guidelines for examining metastasis evolve, and some potentially metastatic lesions can be missed.

\begin{figure}[H]
\centering
%\captionsetup{aboveskip=0pt}
% ^^^^^^^^
\begin{subfigure}[b]{0.42\columnwidth}
%\vspace*{\fill}
  \centering
  \includegraphics[width=\columnwidth,height=3.6cm]{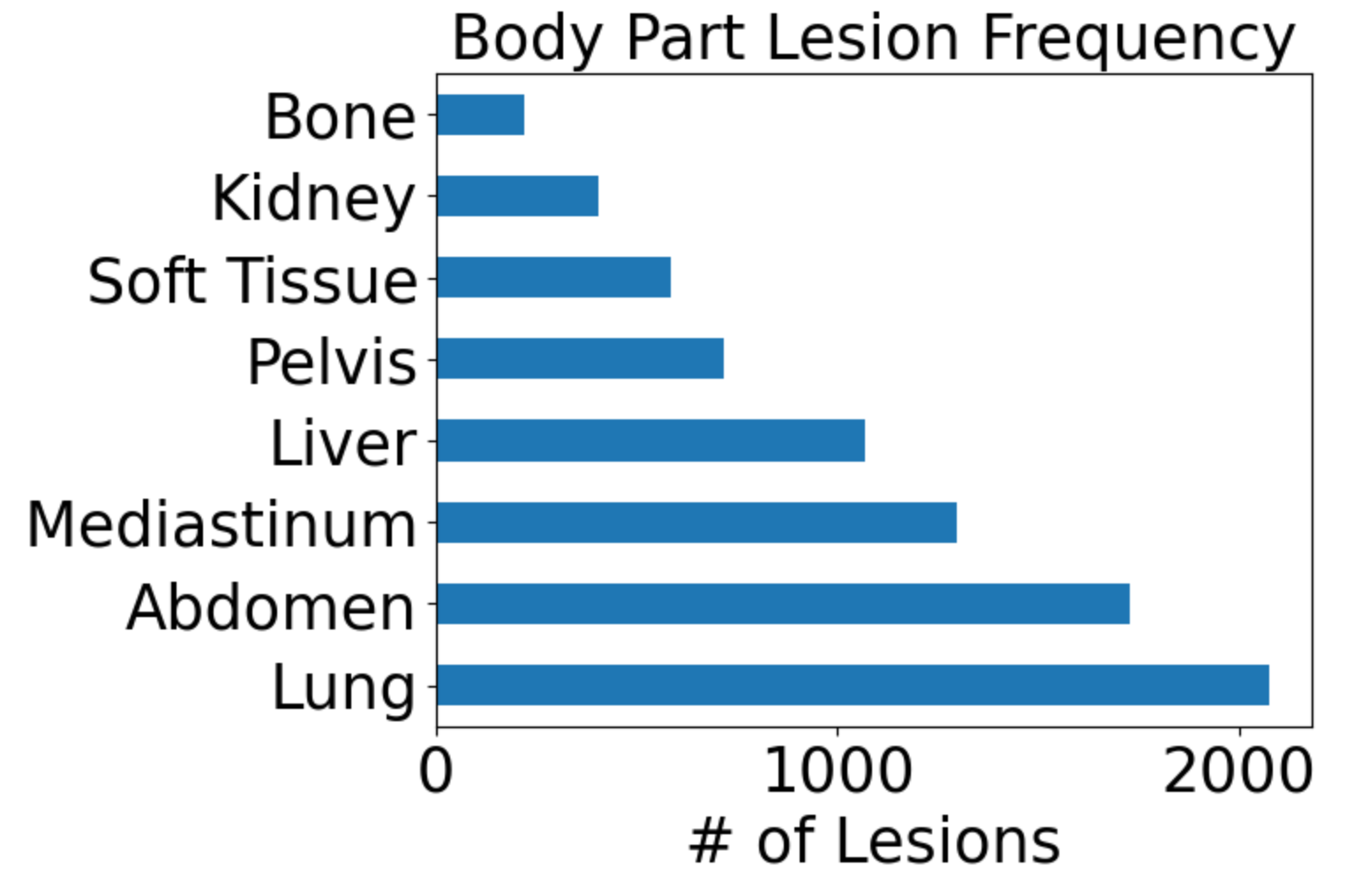}
  \centerline{(a)}
\end{subfigure} 
% ^^^^^^^^
\begin{subfigure}[b]{0.42\columnwidth}
%\vspace*{\fill}
  \centering
  \includegraphics[width=\columnwidth,height=3.6cm]{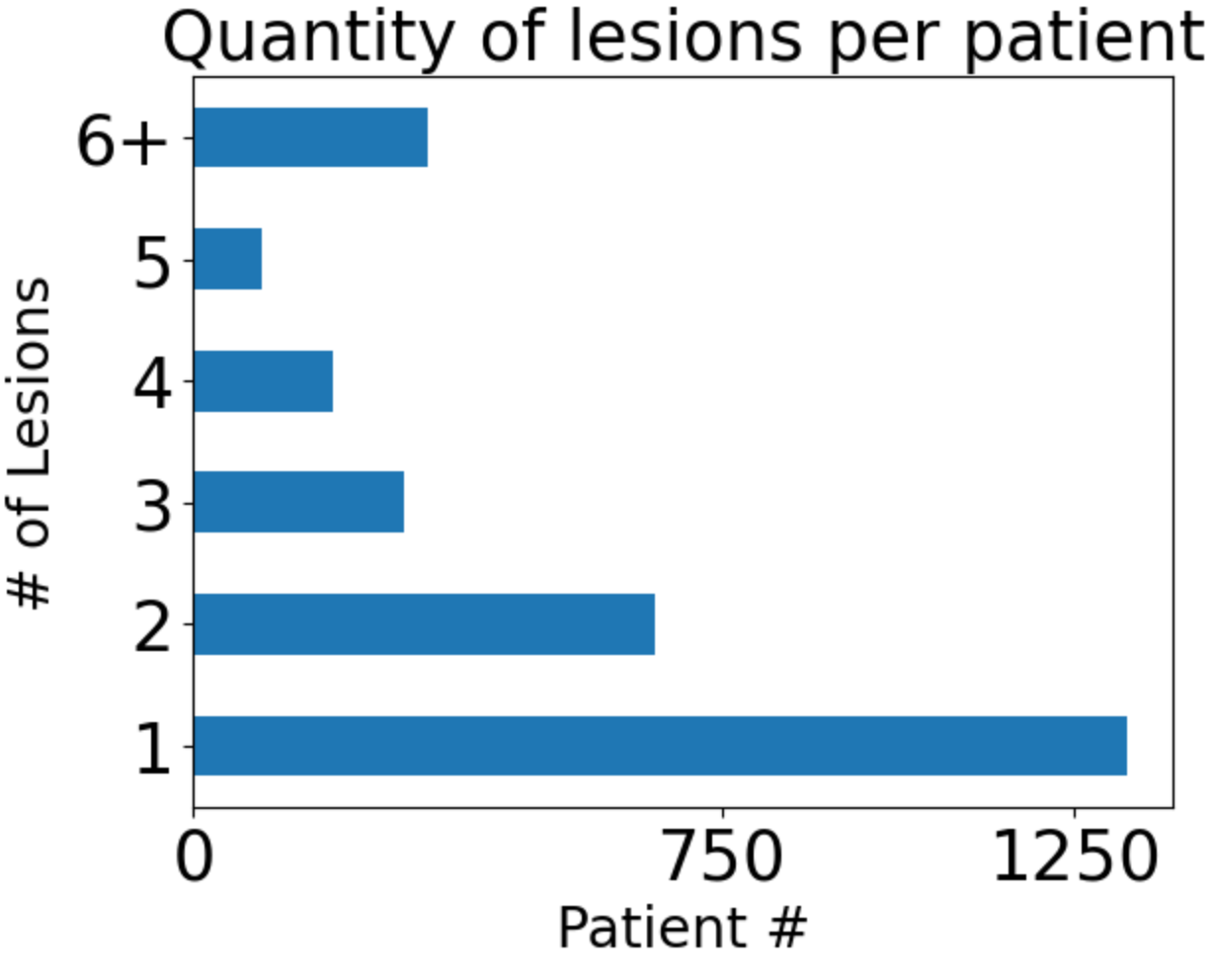}
  \centerline{(b)}
\end{subfigure} 
% ^^^^^^^^
\begin{subfigure}[b]{0.42\columnwidth}
\vspace*{\fill}
  \centering
  \includegraphics[width=\columnwidth,height=3.6cm]{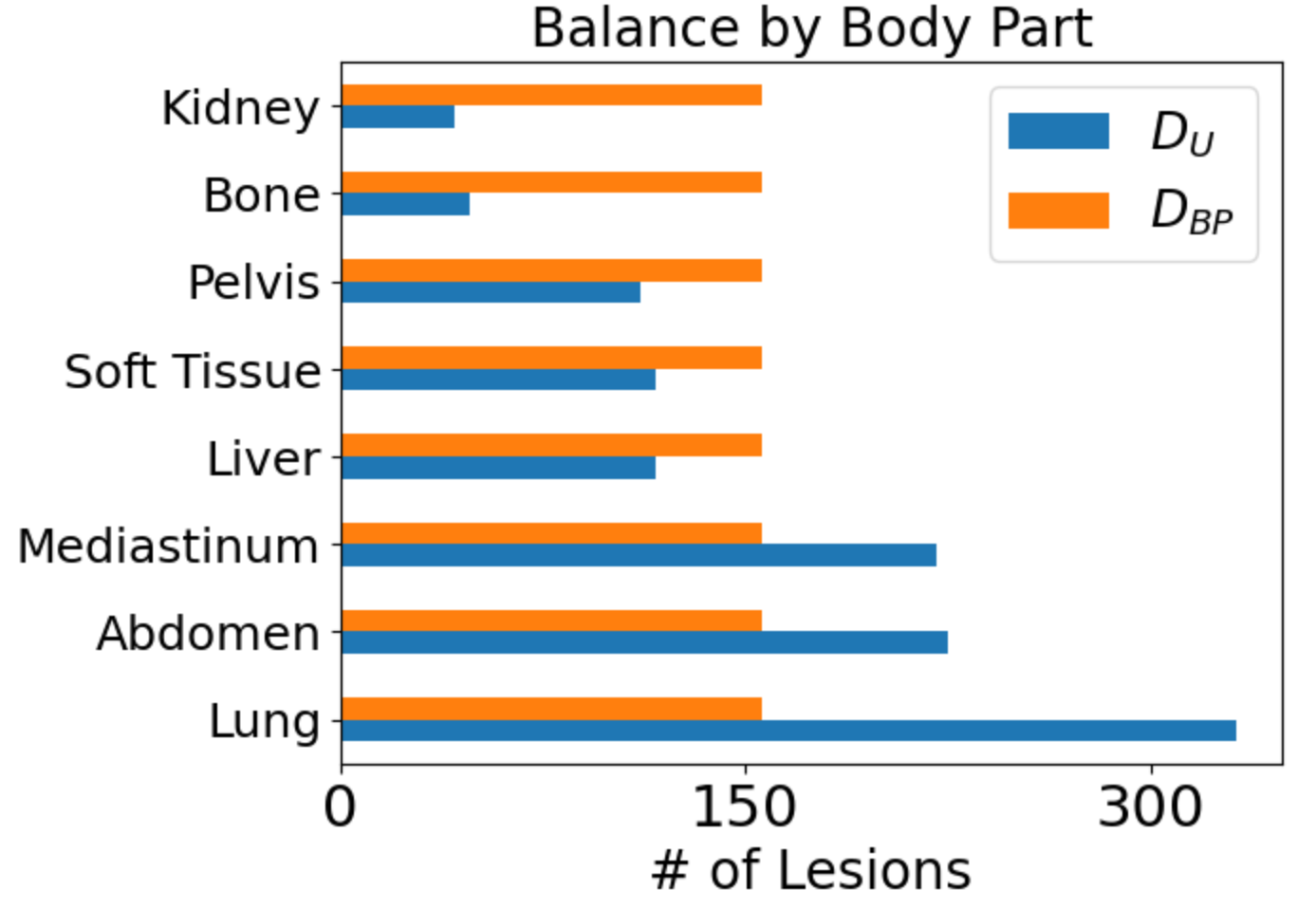}
  \centerline{(c)}
\end{subfigure} 
% ^^^^^^^^
\begin{subfigure}[b]{0.42\columnwidth}
\vspace*{\fill}
  \centering
  \includegraphics[width=\columnwidth,height=3.6cm]{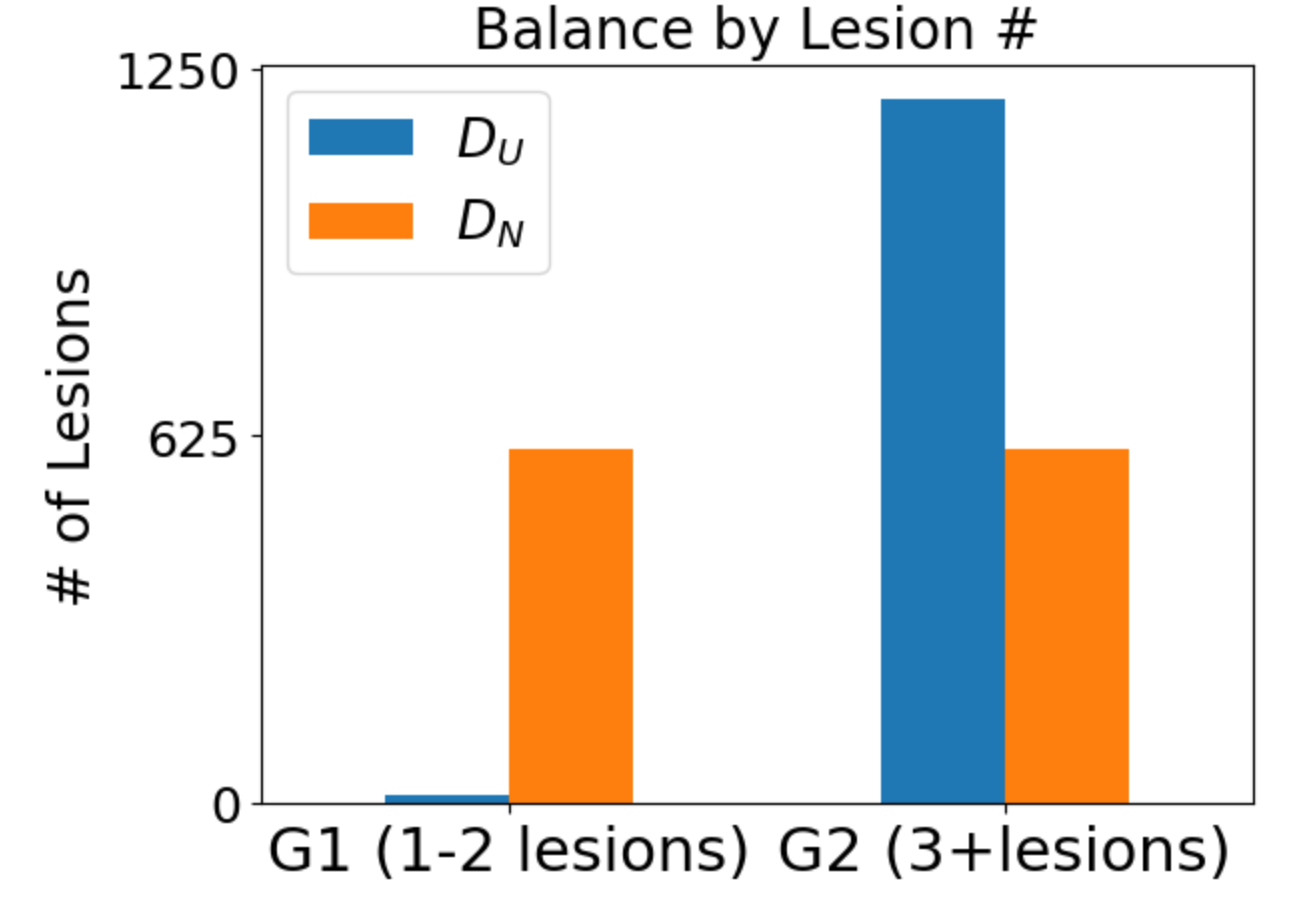}
  \centerline{(d)}
\end{subfigure} 
% ^^^^^^^^
\begin{subfigure}[b]{0.4\columnwidth}
\vspace*{\fill}
  \centering
  \includegraphics[width=\columnwidth,height=3.45cm]{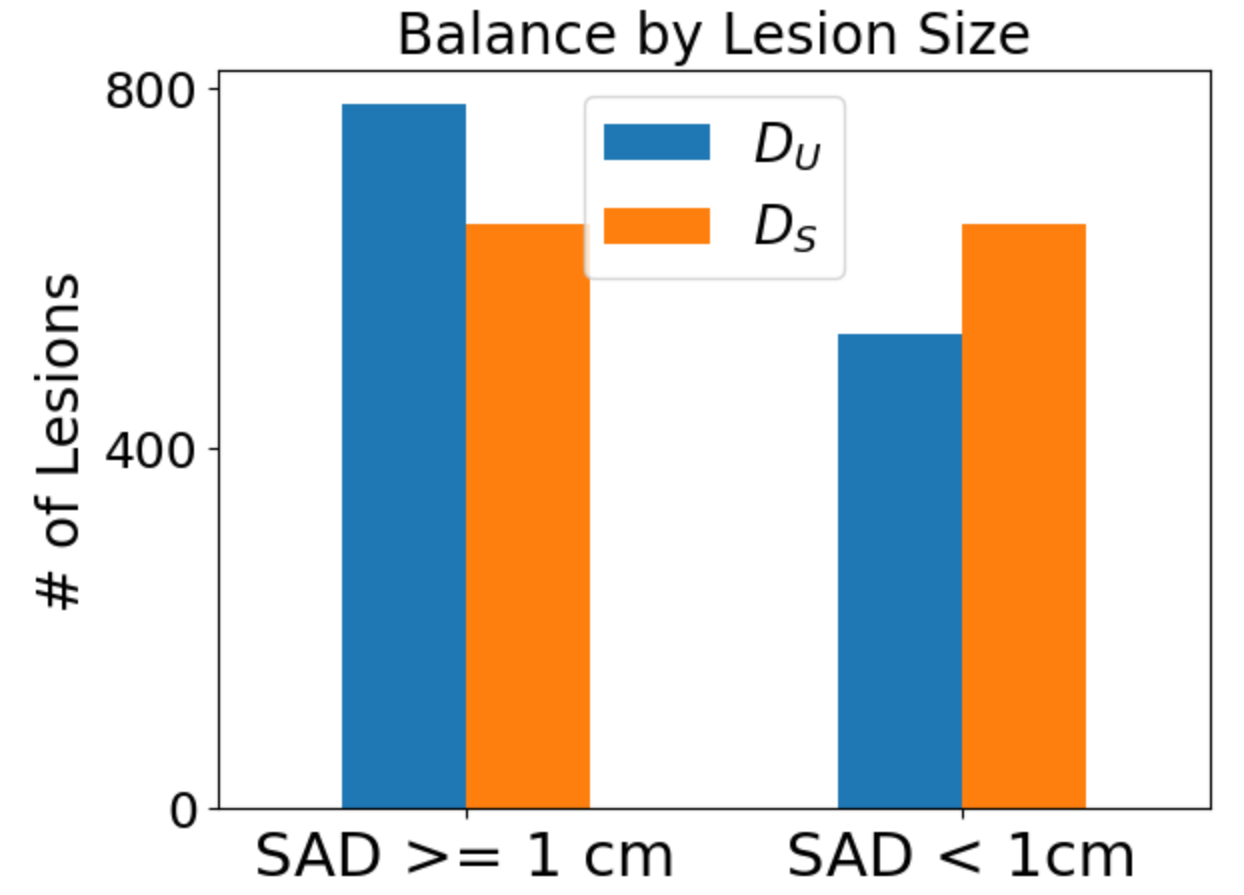}
  \centerline{(e)}
\end{subfigure} 
% ^^^^^^^^
\begin{subfigure}[b]{0.59\columnwidth}
\vspace*{\fill}
  \centering
  \includegraphics[width=\columnwidth,height=3.35cm]{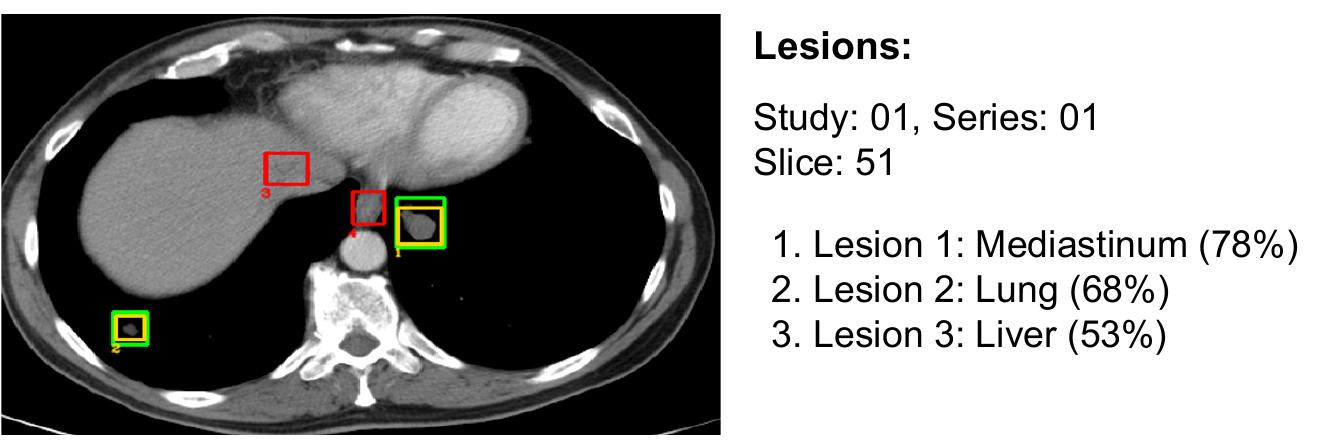}
  \centerline{(f)}
\end{subfigure} 

\caption{(a) shows the lesion distribution per body part label in the DeepLesion dataset \cite{Yan2018_DeepLesion} with certain over-represented and under-represented classes. (b) shows the number of patients with a specific number of lesions annotated. (c) Compared to the unbalanced dataset ${D}_{U}$, our dataset ${D}_{BP}$ balanced the number of lesions across the different body part classes (orange). (d) shows the lesion distribution for patients who were divided into two groups: G1 had patients with 1-2 lesions and G2 had patients with 3+ lesions. Compared to ${D}_{U}$, dataset ${D}_{N}$ (orange) had an equal number of lesions in G1 and G2. The number of patients in each group was not balanced. (e) shows the lesion distribution categorized by the short axis diameter (SAD) length. Compared to ${D}_{U}$, in dataset ${D}_{S}$ the number of lesions with SAD $\geq$ 1cm and SAD $<$ 1cm were balanced (orange). (f) Four lesions were detected in the chest area. Green boxes: GT, yellow boxes: TP, red boxes: FP. The top-3 predictions, their labels, and confidence scores were compiled into a structured ``Lesions'' sub-section for entry into the ``Findings'' section of a radiology report. Only lesions that were predicted with confidences $\geq$50\% are shown. Figure is best viewed electronically in color.
}
\label{fig:data_char}
\end{figure}

Recently, many automated approaches have been proposed for universal lesion detection \cite{Yan2021_LENS,Cai2021_LesionHarvester,Yang2020_AlignShift,Yang2021_A3D,Li2022_SATR,Cai2020_LesionTracker,Tang2022_TformerTracker} on the DeepLesion dataset with state-of-the-art results. The DeepLesion dataset contains 32,735 lesions annotated by radiologists in 32,120 axial CT slices from 10,594 studies of 4,427 patients. The dataset is divided into 70$\%$ train, 15$\%$ validation, and 15$\%$ test splits respectively. Eight (8) lesion-level tags (bone, abdomen, mediastinum, liver, lung, kidney, soft tissue, and pelvis) are available for only the validation and test splits. Prior works utilize the entire dataset for development and testing, while only a handful have gone beyond lesion detection and addressed clinical issues \cite{Cai2020_LesionTracker,Tang2022_TformerTracker,Yan2019_lesa,Yan2019_MULAN}. However, as shown in Fig. \ref{fig:data_char}(a), there is a severe class imbalance in the DeepLesion dataset with over-representation of certain classes (lung, abdomen, mediastinum, and liver) in contrast to other under-represented classes (pelvis, soft tissue, kidney, bone). This imbalance has not been addressed in prior work; e.g., in \cite{Yan2021_LENS}, public datasets for lung nodules (LUNA dataset \cite{Setio2017_LUNA}), liver tumors (LITS dataset\cite{Bilic2019_LITS}), and lymph nodes (NIH Lymph Node dataset \cite{Roth2014_NIHLN}) were added to improve detection. However, this solely increased the data quantities (and detection performance) of the over-represented classes without affecting the under-represented classes. Tackling class imbalances has potential clinical implications, such as improving interval change detection (lesion tracking over time) \cite{Cai2020_LesionTracker,Tang2022_TformerTracker}. 

In this paper, we addressed the class imbalance in the DeepLesion dataset by using only the annotated subset (30$\%$) to train a state-of-the-art VFNet model \cite{Zhang2021_vfnet} for lesion detection and classification. In a limited data-driven manner, we conducted experiments that balanced the training data according to: 1) the body part that the lesion was identified in, 2) the number of lesions observed in a patient, and 3) the size of the lesions. Through balancing the data by the body part label, we have shown a consistent increase in detection sensitivity for under-represented (UR) classes along with a minimal sensitivity drop for over-represented (OR) classes. This trend was also seen with other detectors, such as Faster RCNN \cite{Ren2017_fasterrcnn}, RetinaNet \cite{Lin2017_retinanet}, and FoveaBox \cite{Kong2019_foveabox}. Additionally, we saw recalls for all classes improve with the VFNet model through our experiment that balanced the lesions according to their size. Moreover, we provide a structured reporting guideline by creating a dedicated ``Lesions'' sub-section for entry into the ``Findings'' section of a radiology report. The ``Lesions'' sub-section contains a structured list of detected lesions along with their body part tags, detection confidence, and series and slice numbers. To the best of our knowledge, we are the first to show a class imbalance in the DeepLesion dataset and have taken data-driven steps to address it in the context of lesion detection and classification.

% b - There were many patients with only 1-2 lesions in the study.
% c - the unbalanced dataset ${D}_{U}$ (blue), there are more lesions for certain classes than others.
% d - In ${D}_{U}$, G2 contained more lesions in comparison to G1.
% In ${D}_{U}$ (blue), there are more lesions with SAD $\geq$1cm in contrast to SAD $<$1cm.

% ================================================
% ================================================
\section{Methods}
\label{methods}
% ================================================

In this section, we briefly describe the neural networks that were employed for lesion detection and tagging. Our goal is to improve an existing model's robustness against class imbalances using data-driven approaches. 

% We also elaborate on the mechanism for combining predictions from various epochs of a trained model.

\smallskip
\noindent
\textbf{State-of-the-Art Detectors} Various state-of-the-art detectors were employed for lesion detection and tagging in CT slices. Notably, we used: 1) VFNet \cite{Zhang2021_vfnet}, 2) Faster RCNN \cite{Ren2017_fasterrcnn}, 3) RetinaNet \cite{Lin2017_retinanet}, and 4) FoveaBox \cite{Kong2019_foveabox}. Faster RCNN is a two-stage anchor-based detector in which region proposals for regions-of-interest (ROI) were generated by the first stage, followed by a second stage that classified these proposals and regressed the bounding box coordinates. RetinaNet\cite{Lin2017_retinanet} is an anchor-free detector that utilized the focal loss to solve a common class imbalance problem in detection, wherein proposals were sampled in non-informative ROIs of the image instead of salient object locations. FoveaBox used a ResNet-50 backbone to generate feature maps from the input and a fovea head network that estimated the coordinates in an image that may be potentially covered by an object ROI. Finally, VFNet combined a Fully Convolutional One-Stage Object (FCOS) detector \cite{Tian2019_fcos} (without the centerness branch), an Adaptive Training Sample Selection (ATSS) mechanism \cite{Zhang2019_ATSS}, which selected high quality ROI candidates during training, and a novel IoU-aware varifocal loss \cite{Zhang2021_vfnet} to detect ROI. After model training was completed, Weighted Boxes Fusion (WBF) \cite{Solovyev2021} was used to combine the numerous predictions and improve the precision/recall metric. Supplementary material contains implementation details for the models.

%\noindent
%\textbf{Weighted Boxes Fusion (WBF).} At test time, multiple epochs from either a single run or multiple runs can be used for prediction and tagging of lesions. However, numerous predictions can be obtained from each epoch, and most of these predictions cluster together in common areas in the image. As many of the predictions can be false positives, they can significantly reduce the overall precision and recall measures for detection. To circumvent this problem, we used weighted boxes fusion (WBF) to consolidate the clusters into one, thereby diminishing the false positive counts.

% ================================================

% ================================================
\section{Experiments}
\label{experiments}
% ================================================

\noindent
\textbf{Dataset.} The NIH DeepLesion dataset \cite{Yan2018_DeepLesion} contains keyslices that were annotated with 1-3 lesions per slice and 30mm of context above and below the keyslice was provided. Annotations were made using RECIST measurements \cite{Schwartz2016}, from which 2D bounding boxes were extracted for each lesion. Eight (8) lesion-level tags (bone - 1, abdomen - 2, mediastinum - 3, liver - 4, lung - 5, kidney - 6, soft tissue - 7, and pelvis - 8) were available for only the validation and test splits. The lesion tags were obtained through a body part regressor \cite{Yan2018_BPR}, which provided a continuous score that represented the normalized position of the body part for a slice in a CT volume (e.g., liver, lung, kidney etc.). The body part label for the slice was assigned to any lesion annotated in that slice \cite{Yan2018_DeepLesion,Yan2018_BPR}. As the DeepLesion dataset contained multiple visits of the same patient, only lesions from the first patient visit were kept \cite{Cai2020_LesionTracker} to maintain uniqueness during training. This process left 26,034 lesions from 25,568 slices in the dataset. Next, we removed lesions that did not contain a body part label (the training split) leaving us with a limited dataset ${D}_{L}$ containing 8,104 lesions from 7,953 slices. ${D}_{L}$ contained $\sim$24.75\% of the original DeepLesion dataset, and was split into 60\% training, 20\% validation, 20\% testing splits with unique patients in each split. This 60\% training split was still unbalanced and in our experiments (see below), we utilized only $\sim$6\% of DeepLesion (1331 lesions, 1309 slices).

% HOW I DID THE DATA: 1st, I removed repeat lesions (lesions that were previously identified using longitudinal lesion tracker from the cai et al paper), this left 26,034 lesions and 25,568 slices. From this, I filtered out lesions that did not contain a body part label, so that all of our data was annotated (8,104 lesions and 7,953 slices). From this new dataset, I split the patients into 60, 20, 20 splits. From the 60% patient training split, I performed various forms of balancing by lesion quantity. 

\smallskip
%\medskip
\noindent
\textbf{Experimental Design.} The unbalanced lesion distribution per body part label in Fig. \ref{fig:data_char}(a) and distribution of lesion quantities per patient in Fig. \ref{fig:data_char}(b) led us to design four experiments with a limited annotated dataset ${D}_{L}$. In the first experiment ${E}_{BP}$, we generated a dataset balanced by body part label ${D}_{BP}$; As seen in Figs. \ref{fig:data_char}(a) and \ref{fig:data_char}(c), the body part label with the lowest data quantity (``Bone'') was identified and the data quantities in the remaining labels were matched to the lowest quantity. The intent was to emphasize that all lesion classes were equally likely during training through sample selection. In the second experiment ${E}_{N}$, we created a dataset balanced by the number of lesions ${D}_{N}$ any given patient had. From Fig. \ref{fig:data_char}(b), there are a large number of patients with 1-2 lesions and fewer patients with 3+ lesions. For ${E}_{N}$, we first created two groups (G1 and G2) and sampled patients for each group such that each group contained the same number of lesions as shown in Fig. \ref{fig:data_char}(d). The aim was to provide a balanced number of lesions per patient such that the model witnessed patients with varying number of lesions at test time with equal likelihood. Our third experiment ${E}_{S}$ was clinically oriented, and we produced a dataset balanced by the lesion size ${D}_{S}$. Lesions with SAD $\geq$ 10mm were collected in one group while those with SAD $<$ 10mm were present in the second group. The objective was to create a dataset with equal numbers of lesions divided according to their size as both smaller and larger lesions are equally likely at test time. In our fourth and final experiment ${E}_{U}$, we generated an unbalanced dataset ${D}_{U}$ with a random sample of the training split of ${D}_{L}$, such that it had similar distributions (random) of labels as shown in Figs. \ref{fig:data_char}(c)-(e). 

% ================================================
\section{Results and Discussion}
\label{resultsAndDiscussion}
\begin{figure}[t]
\centering
%\captionsetup{aboveskip=0pt}
% ^^^^^^^^
\begin{subfigure}[b]{0.242\columnwidth}
%\vspace*{\fill}
  \centering
  \includegraphics[width=\columnwidth,height=2cm]{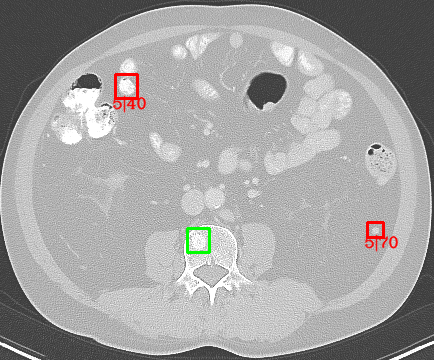}
  \includegraphics[width=\columnwidth,height=2cm]{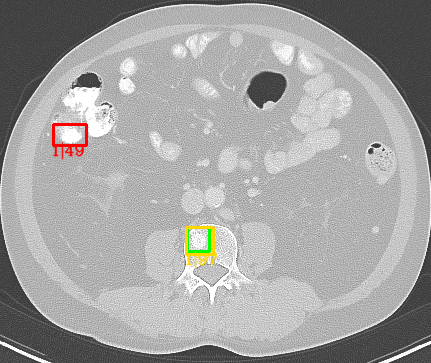}
  \includegraphics[width=\columnwidth,height=2cm]{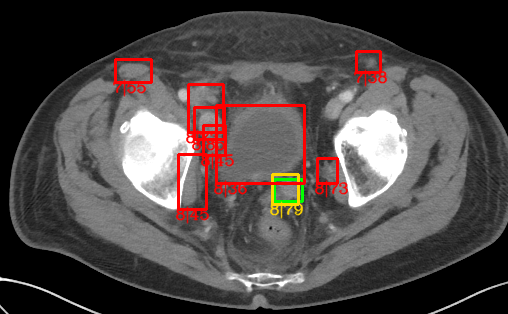}
  \includegraphics[width=\columnwidth,height=2cm]{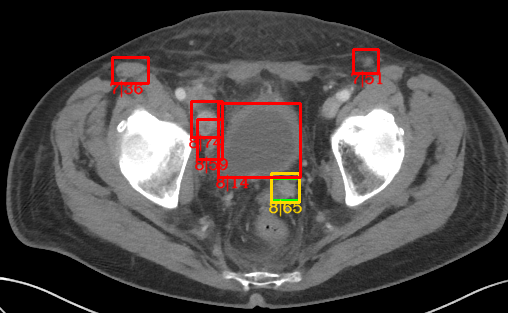}
  \centerline{(a) VFNet}
\end{subfigure} 
% ^^^^^^^^
\begin{subfigure}[b]{0.242\columnwidth}
%\vspace*{\fill}
  \centering
  \includegraphics[width=\columnwidth,height=2cm]{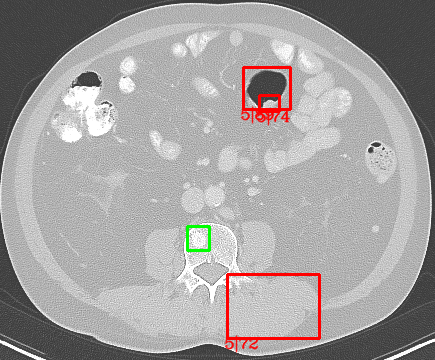}
  \includegraphics[width=\columnwidth,height=2cm]{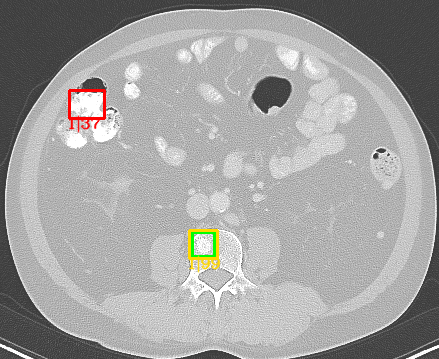}
  \includegraphics[width=\columnwidth,height=2cm]{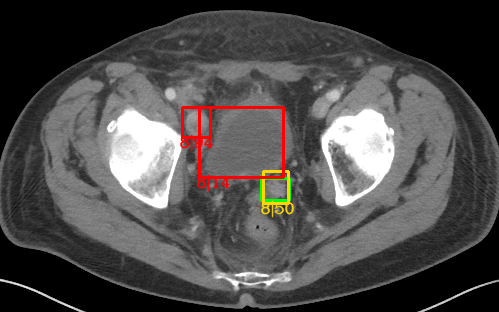}
  \includegraphics[width=\columnwidth,height=2cm]{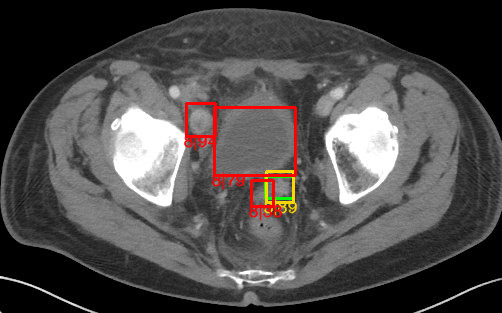}
  \centerline{(b) Faster RCNN}
\end{subfigure} 
% ^^^^^^^^
\begin{subfigure}[b]{0.242\columnwidth}
%\vspace*{\fill}
  \centering
  \includegraphics[width=\columnwidth,height=2cm]{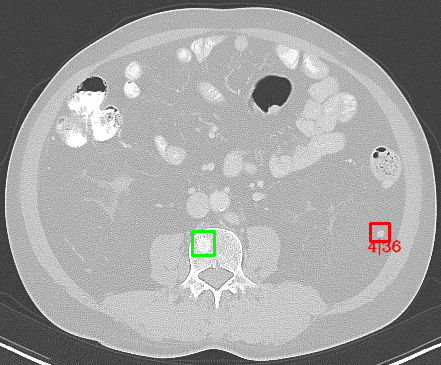}
  \includegraphics[width=\columnwidth,height=2cm]{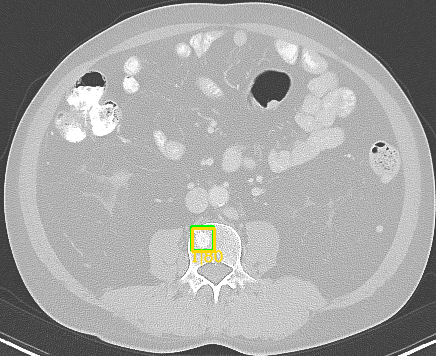}
  \includegraphics[width=\columnwidth,height=2cm]{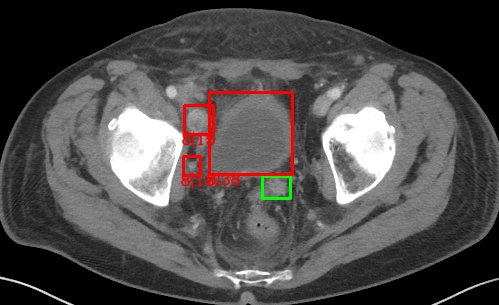}
  \includegraphics[width=\columnwidth,height=2cm]{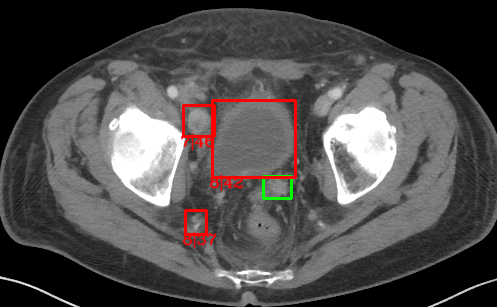}
  \centerline{(c) FoveaBox}
\end{subfigure} 
% ^^^^^^^^
\begin{subfigure}[b]{0.242\columnwidth}
%\vspace*{\fill}
  \centering
  \includegraphics[width=\columnwidth,height=2cm]{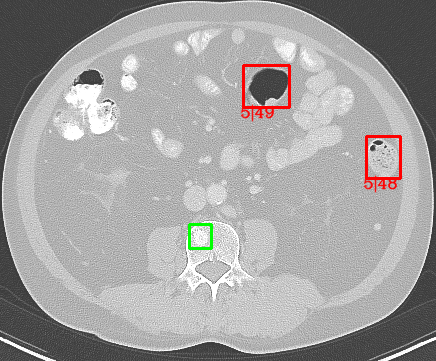}
  \includegraphics[width=\columnwidth,height=2cm]{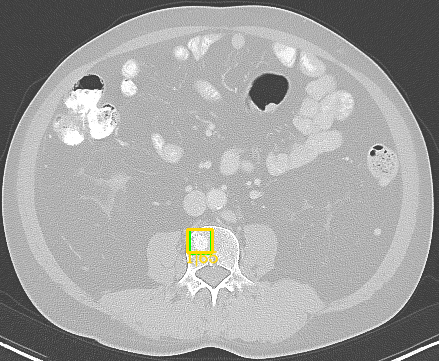}
  \includegraphics[width=\columnwidth,height=2cm]{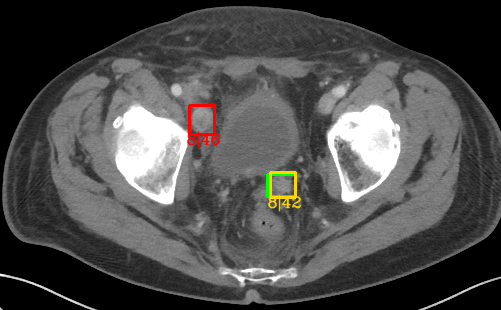}
  \includegraphics[width=\columnwidth,height=2cm]{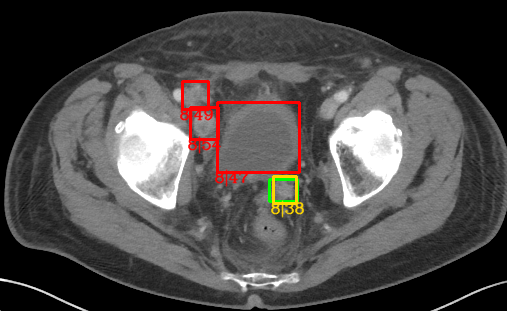}
  \centerline{(d) RetinaNet}
\end{subfigure} 
% ^^^^^^^^
\caption{ Columns (a)-(d) show outputs of the various models on slices from CT volumes of two different patients. The first row of each pairing represents the model output after being trained on an unbalanced ${D}_{U}$ dataset, while the second row shows results when trained on a dataset balanced by body part labels ${D}_{BP}$. Green boxes: GT, yellow boxes: TP, red boxes: FP. The predicted classes and confidence scores are also shown.  The first pair shows that models models trained with ${D}_{U}$ did not identify and classify a ``Bone'' lesion correctly (first row), whereas one trained on ${D}_{BP}$ did (second row). Particularly, VFNet trained on ${D}_{BP}$ predicted correctly with a confidence on 97\%. The second pair shows fewer FP for VFNet with ${D}_{BP}$, and a missed detection for FoveaBox (last row). }
\label{fig:qual_images}
\end{figure}

\textbf{Results.} Table \ref{results_SAD_above_1cm} and Fig. \ref{fig:qual_images} display the results of our experiments at 4 FP and 30\% IoU overlap \cite{Mattikalli2022} on lesions with SAD $\geq$ 1cm, which are generally more clinically significant lesions. Table 1 in the supplementary material reflects our experimental results on lesions with SAD < 1cm. In contrast to the experiment with unbalanced data ${D}_{U}$, our experiment balancing body part labels ${E}_{BP}$ improved recalls for 4/4 under-represented (UR) classes (Bone: 80\% vs. 46\%, Kidney: 77\% vs. 61\%, Soft Tissue: 70\% vs. 60\%, Pelvis: 83\% vs. 76\%) across all the models tested. These results are evident for both SAD $\geq$ and < 1cm. Among the over-represented (OR) classes, we see consistent improvements for the ``Lung'' category across all models and for ``Mediastinum'' label for all models except FoveaBox. However, the ``Abdomen'' and ``Liver'' categories show a decrease in sensitivity across all models. The sensitivity reduction is expected as the number of training samples used for OR classes have been reduced as shown in Fig. \ref{fig:data_char}(c). Although to understand this phenomenon better, we calculated the confusion matrices for each experiment (see supplementary material) using the VFNet model. From the DeepLesion dataset description \cite{Yan2018_DeepLesion}, the ``Soft Tissue'' class encompassed lesions found in the muscle, skin, or fat, while the ``Abdomen'' class was a ``catch-all'' term for all abdominal lesions that were not ``Kidney'' or ``Liver'' masses. Anatomically however, ``Kidney'' and ``Liver'' are organs in close proximity to one another and axial slices often show cross-sections of both organs in the same slice. This is reflected in the confusion matrix as the ``Abdomen'', ``Kidney'' and ``Liver'' labels are confused with each other most often. Comparing ${E}_{U}$ with ${E}_{N}$ (balancing by lesion number), recalls improved for only 1/4 UR classes (Kidney) for Faster RCNN and FoveaBox, and 2/4 UR classes for RetinaNet (Kidney and Soft Tissue) and VFNet (Kidney, soft tissue) respectively. For the OR classes, only ``Liver`` improved for VFNet, 2/4 classes improved for Faster RCNN and FoveaBox (Abdomen, Lung) respectively, and 3/4 classes improved for RetinaNet (Mediastinum, Lung, Liver). Compared against ${E}_{BP}$, recalls were lower for all UR classes except for VFNet, which did well on 2/4 classes (Soft Tissue and Pelvis). For the OR classes, 2/4 classes improved for Faster RCNN (Abdomen, Liver), 3/4 classes improved for VFNet and RetinaNet (Abdomen, Mediastinum and Liver), and all 4 classes improved for FoveaBox. 

% ^^^^^^^^
% ${D}_{U}$
% ${D}_{BP}$
% ${D}_{N}$
% ${D}_{S}$
\begin{table}[h]
\centering\fontsize{9}{12}\selectfont % to make font size 9 pt
\setlength\aboverulesep{0pt}\setlength\belowrulesep{0pt} % intersect vert and horiz lines
\setlength{\tabcolsep}{7pt} % set small spacing between entries of column (default 6pt)
\setcellgapes{3pt}\makegapedcells % small space between row entries (default 3pt)
\caption{Detection sensitivities of different detectors based on different experiments are shown @ 4 FP and an IOU of 0.3 for lesions with a SAD $\geq$ 1cm.}
\begin{adjustbox}{max width=\textwidth}
\begin{tabular}{@{} c|c|c|c|c|c|c|c|c @{}} % @ property can be modified
\toprule
Experiment                                              & Bone      & Kidney  & Soft Tissue      & Pelvis  & Abdomen     & Mediastinum      & Lung       & Liver              \\
\midrule
${E}_{U}$ - Faster R-CNN \cite{Ren2017_fasterrcnn}              & 23.3    & 40.5     & 50.4     & 67.5  & 57.7    & 79.6     & 66.8     & \textbf{77.3}    \\
${E}_{BP}$ - Faster R-CNN \cite{Ren2017_fasterrcnn}             & \textbf{63.3}      & \textbf{75.1}     & \textbf{58.1}  & \textbf{68.5} & 55.6     & \textbf{83.3}     & \textbf{74.9}    & 69.8   \\
${E}_{N}$ - Faster R-CNN \cite{Ren2017_fasterrcnn}        & 16.6  &   49.3  &   44.1  & 63.2   &  \textbf{65.5}  &   78.7  &  72.6 &   76.9  \\
${E}_{S}$ - Faster R-CNN \cite{Ren2017_fasterrcnn}              &  30.0   &  49.7  &  51.7  &  56.4   &  61.1  &  74.8   &  74.5  & 73.1\\
\midrule
${E}_{U}$ - RetinaNet \cite{Lin2017_retinanet}                  & 21.7    & 38.4     & 48.2    & 55.8     & \textbf{70.5}     & 82.8     & 76.1     & 75.9 \\
${E}_{BP}$ - RetinaNet \cite{Lin2017_retinanet}                        & \textbf{66.7}     & \textbf{66.7}     & \textbf{60.3}     & \textbf{59.8}     & 62.3   & 85.3     & \textbf{79.7}    & 71.0 \\
${E}_{N}$ - RetinaNet \cite{Lin2017_retinanet}           &  27.5 &   53.1  &   26.1  &  49.6  &  68.7   &  \textbf{86.2}  &  77.4   &   \textbf{76.5}  \\
${E}_{S}$ - RetinaNet \cite{Lin2017_retinanet}              & 26.2    &  22.4   &  25.0  &  21.1   &  51.9  &  61.7   &   58.8 & 58.2\\
\midrule
${E}_{U}$ - FoveaBox \cite{Kong2019_foveabox}                   & 28.3     & 46.4    & 54.2     & 59.2    & 64.8     & \textbf{88.3 }    & 69.2    & \textbf{76.7} \\
${E}_{BP}$ - FoveaBox \cite{Kong2019_foveabox}                    & \textbf{65.0}     & \textbf{67.9}    & \textbf{66.7}    & \textbf{63.4}   & 56.2     & 84.5     & \textbf{76.9}    & 70.0 \\
${E}_{N}$ - FoveaBox \cite{Kong2019_foveabox}              &  18.3   & 56.5    &  46.9  & 34.1    &  70.1  &  85.1   &  74.7  & 71.5\\
${E}_{S}$ - FoveaBox \cite{Kong2019_foveabox} &  41.66   &  40.9  & 46.3   &   47.6  & \textbf{71.1 } &  86.8 &  75.1  & 74.9 \\
\midrule
${E}_{U}$ - VFNet \cite{Zhang2021_vfnet}                        & 46.7    & 61.6     & 60.0       & 76.0         & 76.8     & 85.6     & 70.8 & 77.7 \\
${E}_{BP}$ - VFNet \cite{Zhang2021_vfnet}                        & \textbf{80.0}     & \textbf{77.6}     & \textbf{70.7}     & 83.4  & 69.5     & 87.7     & 78.9     & 76.3      \\
${E}_{N}$ - VFNet \cite{Zhang2021_vfnet}              &   28.8  &   63.3  &  63.6  &   73.5  &  69.6  &   78.8  &  68.2  & \textbf{91.0} \\
${E}_{S}$ - VFNet \cite{Zhang2021_vfnet}              &  51.6   &  67.0   & 67.3   &  \textbf{87.2}   &  \textbf{82.1}  &   \textbf{89.8}  & \textbf{82.7}   & 82.1\\
\bottomrule
\end{tabular}
\end{adjustbox}
\label{results_SAD_above_1cm}
\end{table}
% ^^^^^^^^

% ^^^^^^^^
% ^^^^^^^^

In contrast to the ${E}_{U}$ experiment, in the ${E}_{S}$ experiment (balancing by lesion size), VFNet recalls were always better across all classes for both SAD $\geq$ 1cm and < 1cm. Only one UR class showed improved recall for RetinaNet and FoveaBox (Bone) respectively, while 3/4 UR classes did better for Faster RCNN (Bone, Kidney, Soft Tissue). In the OR classes, RetinaNet did worse on all classes, and 2/4 OR classes showed improvements for Faster RCNN and FoveaBox (Abdomen, Lung). In contrast to the ${E}_{BP}$ experiment, sensitivities were lower for all UR classes except for the ``Pelvis'' class with VFNet. For the OR classes, RetinaNet did not show improvements for any class, 2/4 classes improved for Faster RCNN (Abdomen, Liver), and 3/4 classes improved for FoveaBox (Abdomen, Mediastinum, Liver). Compared against the ${E}_{N}$ experiment, sensitivities were worse for all UR classes with RetinaNet. They were better for 2/4 UR classes for FoveaBox (Bone and Pelvis), and 3/4 UR classes for Faster RCNN (Bone, Kidney, Soft Tissue). On the OR classes, recall was worse for all classes with RetinaNet, improved for 1/4 OR classes with Faster RCNN (Lung), and 3/4 classes for VFNet (Abdomen, Mediastinum, Lung) and all classes for FoveaBox. 

\noindent
\textbf{Discussion.} In contrast to previous work, we have shown that through effective data exploration of the DeepLesion dataset, the recalls for all models across all the under-represented classes were improved. Specifically, our ${E}_{BP}$ experiment (balancing data by body part labels) displayed this clear improvement. We also saw an increase in sensitivity for the OR classes ``Lung'' and ``Mediastinum`` with Faster RCNN, RetinaNet and VFNet respectively. The ``Abdomen'' and ``Liver'' classes were confused with each other most often. We contend that the ``Abdomen'' and ``Soft Tissue'' labels were generated through a body part regressor, and are ambiguous and non-specific labels that broadly encompass multiple regions in the abdomen. In fact, after we asked a radiologist to re-classify a random sample of 100 lesions with the original term ``Abdomen``, we identified multiple lesions that should be assigned new labels such as ``Liver'', ``Pancreas'', ``Spleen'', ``Muscle'', ``Stomach''  etc. There are many other annotated lesions in DeepLesion for whom the assigned labels may change upon manual inspection. In our experiment ${E}_{N}$ (balancing by number of lesions), we did not see a consistent trend of improvement and hypothesize that this is due to not simultaneously balancing the lesions by the body part labels. Balancing the data by both body part labels and number of lesions proved difficult as it was difficult to categorize patients when they had multiple lesions with different labels. In our ${E}_{S}$ experiment (balancing the lesion size), the recalls for all classes improved with the VFNet model. 

%To the best of our knowledge, we are the first to report on the class imbalance in the DeepLesion dataset, and present results on lesion detection and classification based on body part labels.

We were unable to compare against prior works as limited approaches exist to jointly detect and tag lesion \cite{Yan2019_lesa,Yan2019_MULAN}. One approach \cite{Yan2019_MULAN} used a Mask-RCNN model that required segmentation labels, which we did not create in this work. Furthermore, this approach also provided more descriptive tags, which would require a sophisticated ontology derived from radiology reports (unavailable in DeepLesion dataset) to map them to the body part tags used in this work. To circumvent this issue, we implemented other detection models to prove our consistent results. We also present a clinically useful structured reporting guideline by creating a dedicated ``Lesions'' sub-section for entry into the ``Findings'' section of a radiology report. The ``Lesions'' sub-section contains a structured list of detected lesions along with their body part tags, detection confidence, and series and slice numbers. Furthermore, DeepLesion contains both contrast and non-contrast enhanced CT volumes, but the exact phase information is unavailable in the dataset. Thus, we have not been able to balance the data according to the phase of the CT volume, and this is a limitation of our work. For future work, we plan to conduct an experiment that upsamples the classes with low data points, and balance the data by both the body part label and lesion size.

\section{Conclusion}
% ================================================

In this paper, we have shown that the DeepLesion dataset exhibits a severe imbalance in the number of lesions per body part label. It also contains missing annotations and label tags. We have utilized a limited data subset (6\%, 1331 lesions, 1309 slices) to train a VFNet model to detect lesions and tag them. We conducted three experiments to address the class imbalance and have shown a consistent increase in recalls for UR labels through our experiment ${E}_{BP}$ (Bone: 80\% vs. 46\%, Kidney: 77\% vs. 61\%, Soft Tissue: 70\% vs. 60\%, Pelvis: 83\% vs. 76\%) in contrast to ${E}_{U}$. We have also shown that FasterRCNN, RetinaNet, and FoveaBox perform similarly. In addition, we have shown that balancing data by lesion size helped the VFNet model improve recalls for all classes. To our knowledge, we are the first to show a class imbalance in the DeepLesion dataset and have taken data-driven steps to address it in the context of lesion detection and classification.

\noindent
\textbf{Acknowledgements.} This work was supported by the Intramural Research Program of the National Institutes of Health (NIH) Clinical Center.

% ================================================


\begin{thebibliography}{4}
\small

\bibitem{Eisenhauer2009} E. Eisenhauer et al., ``New response evaluation criteria in solid tumours: revised RECIST guideline (version 1.1)'', Eur J Cancer, 45(2), pp. 228--247 (2009).

\bibitem{Schwartz2016} L. Schwartz et al., \enquote{Recist 1.1—update and clarification: From the recist committee}, European J Cancer, 62, pp. 132--137 (2016).

\bibitem{Yan2021_LENS} K. Yan et al., ``Learning from Multiple Datasets with Heterogeneous and Partial Labels for Universal Lesion Detection in CT'', In: IEEE TMI, 40(10), pp. 2759--2770 (2021).

\bibitem{Cai2021_LesionHarvester} J. Cai et al., ``Lesion Harvester: Iteratively Mining Unlabeled Lesions and Hard-Negative Examples at Scale'', In: IEEE TMI, 40(1), pp. 59--70 (2021). 

\bibitem{Yang2020_AlignShift} J. Yang et al., ``AlignShift: Bridging the Gap of Imaging Thickness in 3D Anisotropic Volumes'', In: MICCAI, pp. 562--572 (2020).

\bibitem{Yang2021_A3D} J. Yang  et al., ``Asymmetric 3D Context Fusion for Universal Lesion Detection'', In: MICCAI, (2021).

\bibitem{Li2022_SATR} L. Han et al., ``SATr: Slice Attention with Transformer for Universal Lesion Detection'', In: arXiv, (2022).

\bibitem{Cai2020_LesionTracker} J. Cai et al., ``Deep Lesion Tracker: Monitoring Lesions in 4D Longitudinal Imaging Studies'', In: IEEE CVPR, (2020).

\bibitem{Tang2022_TformerTracker} W. Tang et al., ``Transformer Lesion Tracker'', In: arXiv, (2022).

\bibitem{Yan2019_lesa} K. Yan et al., ``Holistic and Comprehensive Annotation of Clinically Significant Findings on Diverse CT Images: Learning from Radiology Reports and Label Ontology'', In: IEEE CVPR, (2019).

\bibitem{Yan2019_MULAN} K. Yan et al., ``MULAN: Multitask Universal Lesion Analysis Network for Joint Lesion Detection, Tagging, and Segmentation'', In: MICCAI, (2019).

\bibitem{Setio2017_LUNA} A. A. A. Setio et al., “Validation, comparison, and combination of algorithms for automatic detection of pulmonary nodules in computed tomography images: The LUNA16 challenge'', Med. Image Anal., 42, pp. 1-–13 (2017).

\bibitem{Bilic2019_LITS} P. Bilic et al., "The Liver Tumor Segmentation Benchmark (LiTS)'', In: CoRR, (2019). 

\bibitem{Roth2014_NIHLN} H. Roth et al., "A New 2.5D Representation for Lymph Node Detection Using Random Sets of Deep Convolutional Neural Network Observations'', In: MICCAI, (2014). 

\bibitem{Zhang2021_vfnet} H. Zhang et al., ``VarifocalNet: An IoU-Aware Dense Object Detector'', In: IEEE CVPR, pp. 8514--8523 (2021). 

\bibitem{Ren2017_fasterrcnn} S. Ren et al., ``Faster R-CNN: Towards Real-Time Object Detection with Region Proposal Networks'', In: IEEE PAMI, 39(6), pp. 1137--1149 (2017). 

\bibitem{Lin2017_retinanet} T.Y. Lin et al., ``Focal Loss for Dense Object Detection'', In: IEEE ICCV, pp. 2999--3007 (2017).

\bibitem{Kong2019_foveabox} T. Kong et al., ``FoveaBox: Beyond Anchor-based Object Detector'', In: arXiv, (2019). 

\bibitem{Tian2019_fcos} Z. Tian et al., ``FCOS: Fully Convolutional One-Stage Object Detection'', In: IEEE ICCV, pp. 9627--9636 (2019).   

\bibitem{Zhang2019_ATSS} S. Zhang et al., ``Bridging the Gap Between Anchor-based and Anchor-free Detection via Adaptive Training Sample Selection'', In: CoRR, (2019).  

\bibitem{Solovyev2021} R. Solovyev et al., ``Weighted Boxes Fusion: Ensembling Boxes from Different Object Detection Models'', Img. Vis. Comp., 107, (2021). 

\bibitem{Yan2018_DeepLesion} K. Yan et al., \enquote{DeepLesion: automated mining of large-scale lesion annotations and universal lesion detection with deep learning}, J Medical Imaging, 5(3), pp. 1--11 (2018).

\bibitem{Yan2018_BPR} K. Yan et al., \enquote{Unsupervised body part regression via spatially self-ordering convolutional neural networks}, In: IEEE ISBI, pp. 1022--1025, (2018). 

\bibitem{Mattikalli2022} T. Mattikalli et al., ``Universal lesion detection in CT scans using neural network ensembles'', In: SPIE Medical Imaging: Computer-Aided Diagnosis, 12033, (2022).

\end{thebibliography}
\end{document}

\maketitle

\begin{figure}[H]
\centering
%\captionsetup{aboveskip=0pt}
% ^^^^^^^^
\begin{subfigure}[b]{0.325\columnwidth}
\vspace*{\fill}
  \centering
  \includegraphics[width=\columnwidth,height=3.5cm]{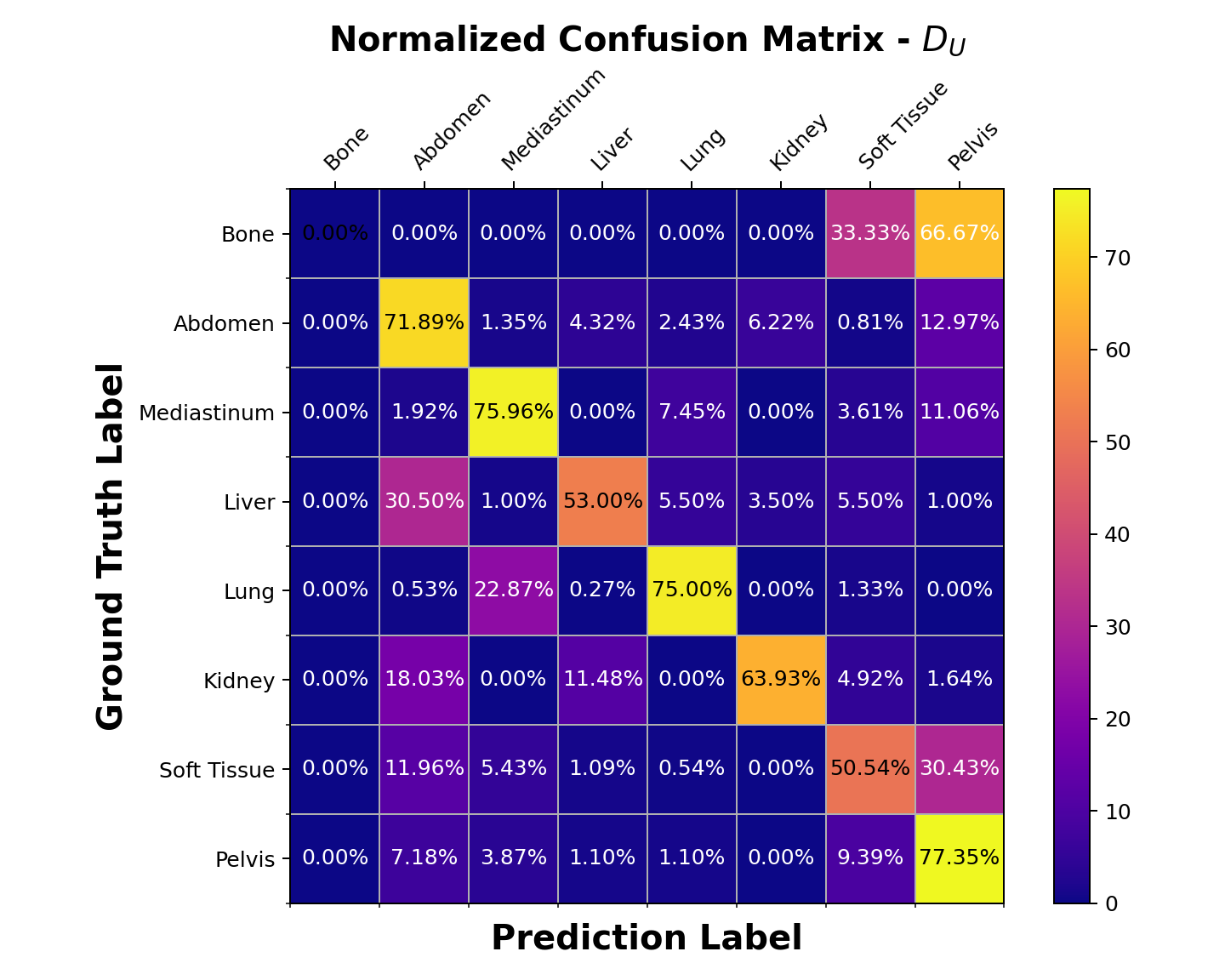}
  \centerline{(a)} 
  \includegraphics[width=\columnwidth,height=3.5cm]{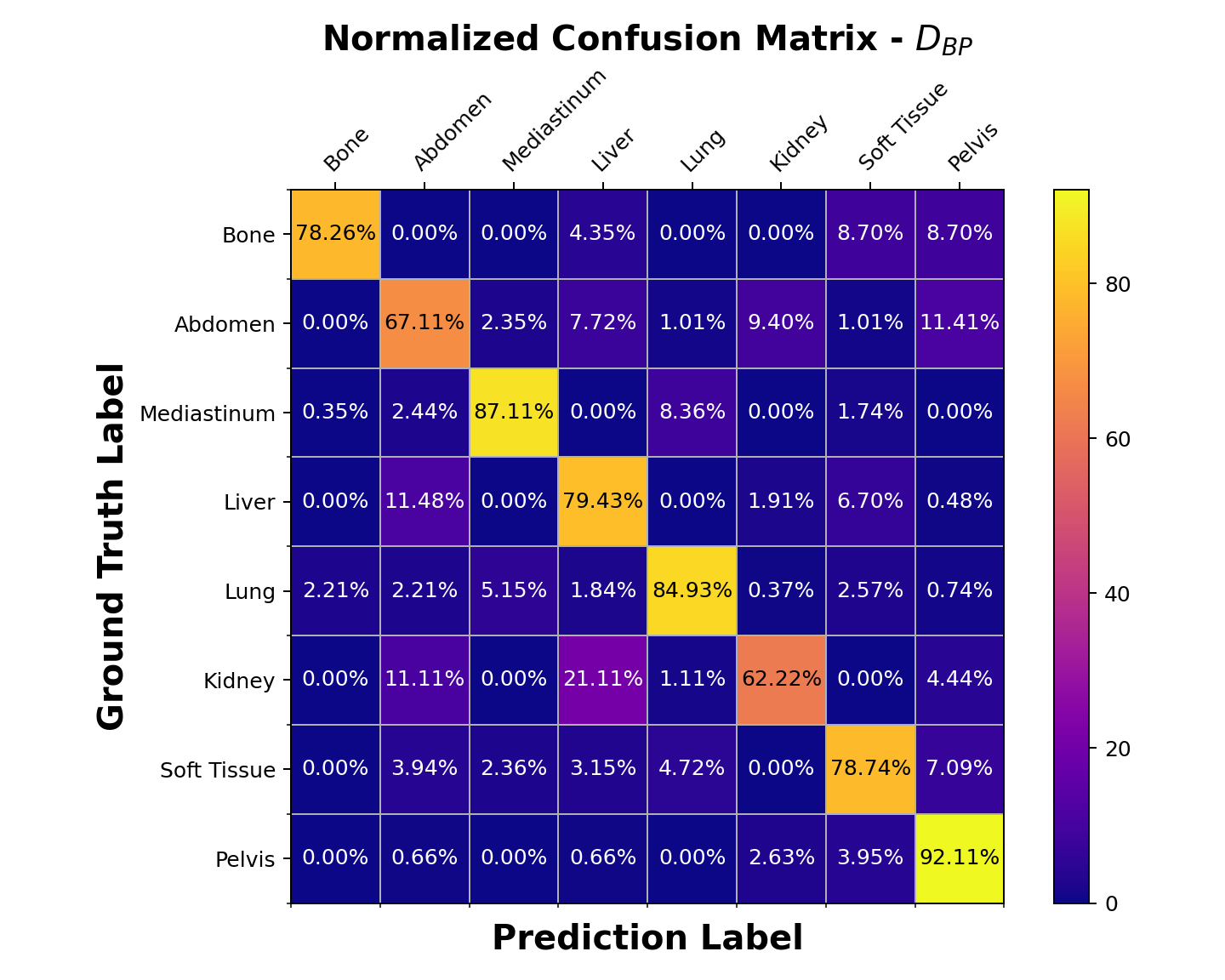}
  \centerline{(b)} 
\end{subfigure} 
% ^^^^^^^^
\begin{subfigure}[b]{0.325\columnwidth}
\vspace*{\fill}
  \centering
  \includegraphics[width=\columnwidth,height=3.5cm]{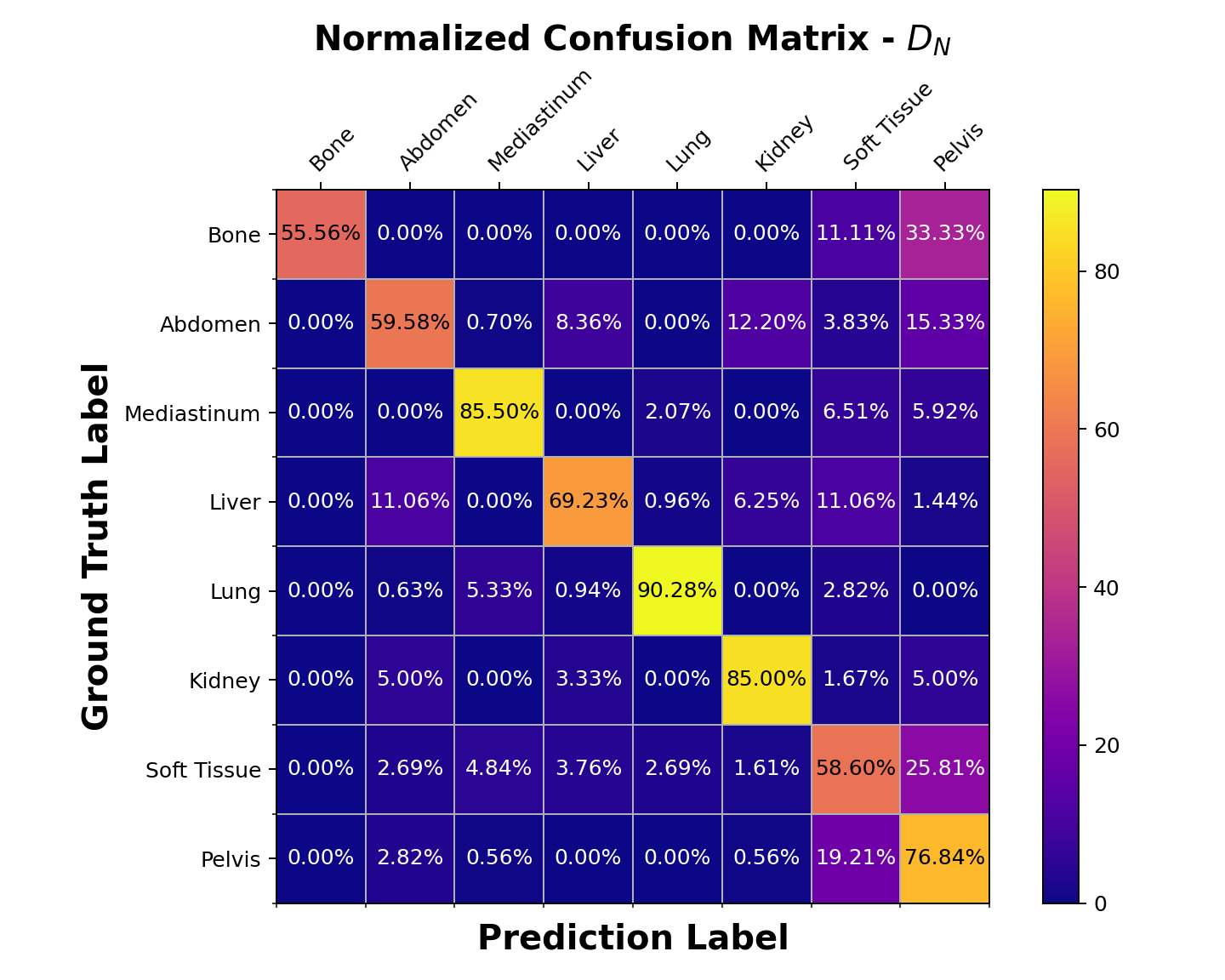}
  \centerline{(c)} 
  \includegraphics[width=\columnwidth,height=3.5cm]{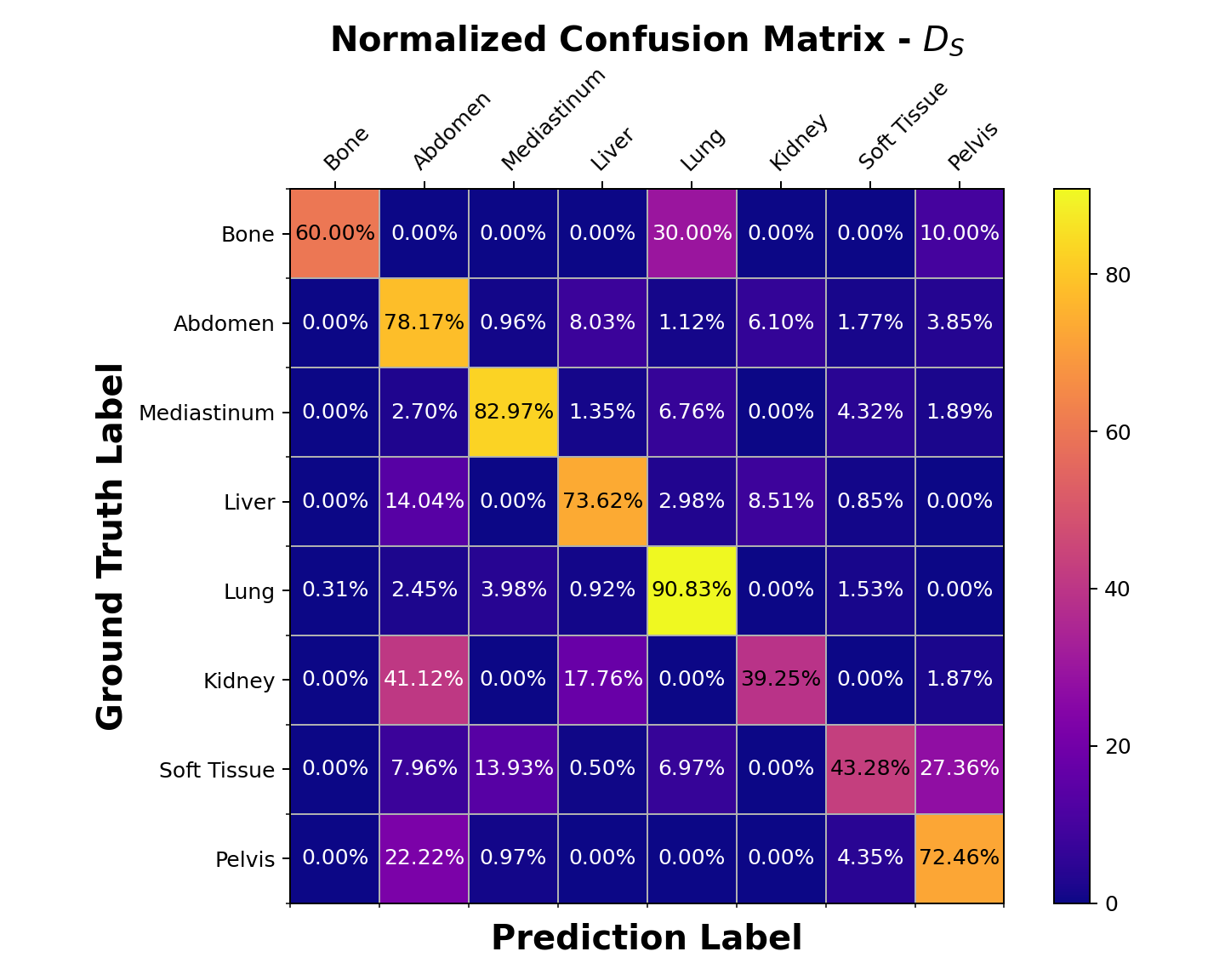}
  \centerline{(d)} 
\end{subfigure} 
% ^^^^^^^^
\begin{subfigure}[b]{0.325\columnwidth}
\vspace*{\fill}
  \centering
  \includegraphics[width=0.75\columnwidth,height=3cm]{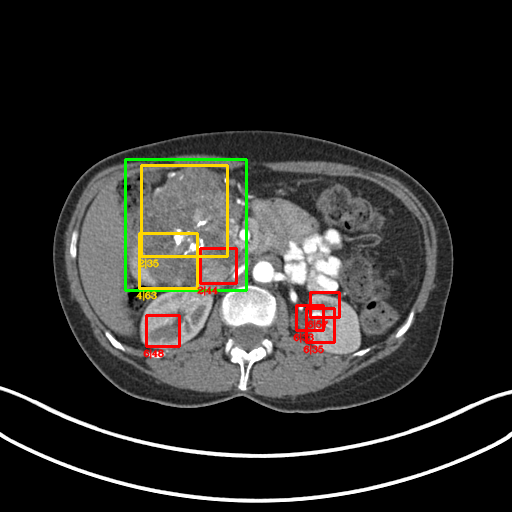}
  \centerline{(e)} 
  \includegraphics[width=0.75\columnwidth,height=3cm]{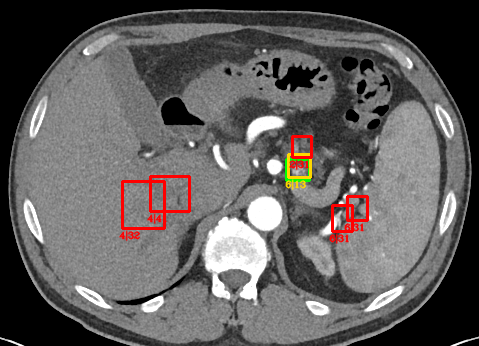}
  \centerline{(f)} 
\end{subfigure} 
% ^^^^^^^^
\caption{(a) Confusion matrix for the experiment $E_{U}$. (b) Confusion matrix for $E_{BP}$. (c) Confusion matrix for $E_{N}$. (d) Confusion matrix for $E_{S}$. (e) Incorrect prediction of a ``Liver'' lesion (yellow, label 4) when GT was ``Abdomen'' (green, label 2). Notice the proximity of the incorrect prediction to the liver region next to it. (f) Incorrect prediction of a ``kidney'' lesion (yellow, label 6) when GT was ``Abdomen'' (green, label 2). Notice the incorrect prediction residing in the pancreo-splenic region.}
\label{fig:staticvsdescending}
\end{figure}
% ^^^^^^^^

\noindent
\textbf{Implementation.} 
The window center and width provided in DeepLesion \cite{Yan2018_DeepLesion} were used to window the Hounsfield units (HU) in a CT slice and clip them to the [0,255] range. Additionally, slices above and below the annotated keyslice were also windowed and normalized. To mimic the radiologist's approach of scrolling through slices in a CT volume, we constructed a 2.5D image with three consecutive slices (middle slice was the annotated keyslice) for training the detectors. Each slice was resized to 512$\times$512 pixels. The backbone for all detectors was a ResNet-50 model, and data was augmented through standard strategies, such as random flips, random rotations, random intensity shifts, resizing, random crops etc. After a grid search, we set the batch size to 2, and the learning rate for VFNet and FasterRCNN to ${1e}^{-3}$ and for RetinaNet and FoveaBox to ${2.5e}^{-3}$ respectively. VFNet ran for 24 epochs, while the others ran for 12 epochs. All experiments were done on a workstation running Ubuntu 16.04LTS and containing a NVIDIA Tesla V100 GPU. Three-fold cross-validation was conducted for each model, and fusion of predictions through WBF was done at test time with the top 5 epochs having the lowest validation loss. Consistent with literature, models were evaluated at a 0.3 IOU threshold \cite{Yan2021_LENS}.

\begin{table}[h]
\centering\fontsize{9}{12}\selectfont % to make font size 9 pt
\setlength\aboverulesep{0pt}\setlength\belowrulesep{0pt} % intersect vert and horiz lines
\setlength{\tabcolsep}{7pt} % set small spacing between entries of column (default 6pt)
\setcellgapes{3pt}\makegapedcells % small space between row entries (default 3pt)
\caption{Performance of different detectors. Values are measures of sensitivity at an IOU of 0.3 for lesions with a SAD< 1 cm @ 4 FP.}
\begin{adjustbox}{max width=\textwidth}
\begin{tabular}{@{} c|c|c|c|c|c|c|c|c @{}} % @ property can be modified
\toprule
Method and Dataset                                              & Bone & Kidney       & Soft Tissue      & Pelvis       & Abdomen     & Mediastinum      & Lung       & Liver       \\
\midrule
${D}_{U}$ - Faster R-CNN \cite{Ren2017_fasterrcnn}              & 8.9  & 8.3     & 53.0     & 40.8   & 46.5     & 68.1     & 40.6     & 80.0    \\ 
${D}_{BP}$ - Faster R-CNN \cite{Ren2017_fasterrcnn}              & \textbf{61.1}    & \textbf{26.6}     & \textbf{65.9}    & \textbf{55.3}     & 43.0    & \textbf{74.4}     & 51.5    & 75.6 \\
${D}_{N}$ - Faster R-CNN \cite{Ren2017_fasterrcnn}            & 14.4  &   13.3  &   45.4  &  54.7  &  \textbf{53.6}  &   70.8  &  46.8  &   80.7  \\
${D}_{S}$ - Faster R-CNN \cite{Ren2017_fasterrcnn}              &  31.1   &  16.6   & 51.5   & 47.8    &  52.1  &   66.0  &  \textbf{54.6}  & \textbf{81.3}\\
\midrule
${D}_{U}$ - RetinaNet \cite{Lin2017_retinanet}                  & 16.6    & 28.3     & 48.4   & 45.2     & \textbf{64.7}     & \textbf{83.0}    & 62.5    & \textbf{85.1} \\
${D}_{BP}$ - RetinaNet \cite{Lin2017_retinanet}                        & \textbf{68.8}     & \textbf{48.3}   & \textbf{65.9}     & \textbf{54.7}     & 61.56   & 74.1     & \textbf{67.1}    & 82.6 \\
${D}_{N}$ - RetinaNet \cite{Lin2017_retinanet}              &  15.0   &   20.0  &  28.4  &   41.5  & 62.5  & 78.1   &  65.6  & 84.8 \\
${D}_{S}$ - RetinaNet \cite{Lin2017_retinanet}              &  25.0   &  12.5   & 31.2   &  19.8   & 54.6   & 53.8    &    48.0 & 66.4\\
\midrule
${D}_{U}$ - FoveaBox \cite{Kong2019_foveabox}                   & 16.6     & 10.0     & 50.0     & 38.3     & 56.1   & 78.2    & 52.0    & 78.7 \\
${D}_{BP}$ - FoveaBox \cite{Kong2019_foveabox}                    & \textbf{53.3}     & \textbf{33.3}     & \textbf{70.4}     & \textbf{57.2}   & 48.9   & 73.8    & 57.2    & 73.5 \\
${D}_{N}$ - FoveaBox \cite{Kong2019_foveabox}              & 14.4    &    26.6 &  49.2  &  37.7   &  59.6  & \textbf{79.4}    &  59.9  & 80.5\\
${D}_{S}$ - FoveaBox \cite{Kong2019_foveabox}              &  37.8   &   23.3  &  53.0  &  42.8  &  \textbf{63.2}  &   74.5 &  \textbf{62.5} &\textbf{84.6} \\
\midrule
${D}_{U}$ - VFNet \cite{Zhang2021_vfnet}                        & 34.4    & 35.0    & 62.1        & 62.8        & 70.1    & 81.5    & 56.2 & 87.9 \\
${D}_{BP}$ - VFNet \cite{Zhang2021_vfnet}                        & \textbf{75.5}     & \textbf{76.6}    & 76.5     & 71.7    & 62.6   & 78.8    & 68.2      & 86.1 \\
${D}_{N}$ - VFNet \cite{Zhang2021_vfnet}              &  23.3  &   65.4 & \textbf{81.7}   & 73.3    &  \textbf{84.5}  & 79.5    & \textbf{76.7}   & 65.4\\
${D}_{S}$ - VFNet \cite{Zhang2021_vfnet}             &   53.3   &  60.0   & 76.5   &   \textbf{85.5}  & 83.8   &  \textbf{87.5}   &  72.9  & \textbf{89.7}\\
\bottomrule
\end{tabular}
\end{adjustbox}
\label{results_SAD_below_1cm}
\end{table}